\documentclass[useAMS,usenatbib,usegraphicx]{mn2e}

\title[The Microlensing Optical Depth Towards the Large Magellanic
      Cloud: Is There A Puzzle?]
      {The Microlensing Optical Depth Towards the Large Magellanic
      Cloud: Is There A Puzzle?}

\author[N.W. Evans \& V. Belokurov]
       {N. Wyn Evans$^1$, Vasily Belokurov$^{1}$\\
       $^1$ Institute of Astronomy, Madingley Rd, Cambridge, CB3 0HA, UK}

\def\FigDir{.}
\def\spose#1{\hbox to 0pt{#1\hss}}
\def\lta{\mathrel{\spose{\lower 3pt\hbox{$\sim$}} \raise
2.0pt\hbox{$<$}}}
\def\gta{\mathrel{\spose{\lower 3pt\hbox{$\sim$}} \raise
2.0pt\hbox{$>$}}}

\begin{document} 

\maketitle

\begin{abstract}
Using neural networks, Belokurov, Evans \& Le Du (2003, 2004) showed
that 7 out of the 29 microlensing candidates towards the Large
Magellanic Cloud (LMC) of the MACHO collaboration are consistent with
blended microlensing and added Gaussian noise. We then estimated the
microlensing optical depth to the LMC to be $0.3 \times 10^{-7} \lta
\tau \lta 0.5 \times 10^{-7}$, lower than the value $\tau =
1.2^{+0.4}_{-0.3} \times 10^{-7}$ claimed by the MACHO collaboration
(Alcock et al. 2000).  There have been independent claims of a low
optical depth to the LMC by the EROS collaboration, who have most
recently reported $ \tau < 0.36 \times 10^{-6}$ (Tisserand et
al. 2006).

Griest \& Thomas (2005) have contested our
calculations. Unfortunately, their paper contains a number of
scientific misrepresentations of our work, which we clarify here. We
stand by our application of the neural networks to microlensing
searches, and believe it to be a technique of great promise. Rather,
the main cause of the disparity between Griest \& Thomas (2005) and
Belokurov et al. (2004) lies in the very different datasets through
which these investigators look for microlensing events. Whilst not
everything is understood about the microlensing datasets towards the
LMC, the latest downward revisions of the optical depth to $(1.0 \pm
0.3) \times 10 ^{-7}$ (Bennett 2005) is within $\lta 2\sigma$ of the
theoretical prediction from stellar populations alone.

Efficiency calculations can correct for the effects of false
negatives, but they cannot correct for the effects of false positives
(variable stars that are mistaken for microlensing).  In our opinion,
the best strategy in a microlensing experiment is to eschew a decision
boundary altogether and so sidestep the vagaries of candidate
selection and efficiency calculations. Rather, each lightcurve should
be assigned a probability that it is a bona fide microlensing event
and the microlensing rate calculated by summing over the probabilities
of all such lightcurves.
\end{abstract}


\section{Introduction}

The microlensing puzzle is: what is the origin of the microlensing
events towards the Large Magellanic Cloud (LMC)? Specifically, what
fraction of the microlensing events are caused by known stellar
components in the Milky Way and by self-lensing of the LMC, and what
fraction by a compact dark matter component in the Milky Way halo.
Griest \& Thomas (2005) argue that there is evidence for an excess
of events above and beyond the contribution of the known stellar
components in the Milky Way and LMC and hence there is evidence for
compact dark objects in the halo.

There are two microlensing collaborations who have heroically
monitored the Magellanic Clouds over many years. They have reported
rather different numbers of events.  After 8 years of monitoring, the
EROS collaboration announced just 3 microlensing candidates towards
the LMC (Lasserre et al. 2000).  By contrast, the MACHO collaboration
(Alcock et al. 1997) first published an analysis of their 2-year
dataset. They found a high microlensing optical depth ($\tau =
2.9^{+1.4}_{-0.9} \times 10 ^{-7}$) based on an 8 event sample.  They
suggested that this was consistent with about $50 \% $ of the halo
within 50 kpc being made of objects with mass $\sim 0.5 M_\odot$. This
optical depth value was superseded by the analysis of 5.7 years of
data, which indicated a somewhat lower optical depth of $\tau =
1.2^{+0.4}_{-0.3} \times 10^{-7}$ based on either 13 or 17 events
(Alcock et al. 2000).

Belokurov, Evans \& Le Du (2004) re-analyzed 22\,000 publicly
available MACHO lightcurves with neural networks, and provided
alternative sets of microlensing events. The subset reanalyzed
contained all the microlensing candidates of Alcock et al. (2000), but
is only a small fraction of the entire public archive of 9 million
MACHO lightcurves We argued that at least some of the events
identified as microlensing by Alcock et al. (2000) may in fact be
contaminants.  We roughly estimated the optical depth as $0.3 \times
10^{-7} \lta \tau \lta 0.5 \times 10^{-7}$ (Evans \& Belokurov 2004)

Subsequently, the EROS collaboration (Tisserand \& Milsztajn 2005)
reported an optical depth to the Large Magellanic Cloud of $\tau =
(0.15 \pm 0.12) \times 10^{-7}$ based on 3 candidates found in 6.7
years of data -- a remarkably low result. Here, the error is
calculated using Han \& Gould's (1995) formula with $\eta = 2.0$. Note
that the EROS estimate provides an upper bound to the contribution of
compact dark halo objects to the total optical depth, as an obvious
disk lens event was removed. Very recently, EROS have reported further
interesting results based on the clever use of a bright subsample of
source stars to minimise contamination (Tisserand et al. 2006). They
find only one microlensing candidate in this subsample and suggest
that the optical depth due to such lenses is $\tau < 0.36 \times
10^{-7}$ at the 95 \% confidence level. If these low values for the
optical depth are accepted, then the stellar populations in the outer
Galaxy and the LMC must provide most of the lenses for the known
events -- as in fact is true in all instances where the location of
the lens can be identified. There are two exotic events towards the
LMC, for which the location of the event can be more or less
inferred. They are the binary caustic crossing event studied by
Bennett et al. (1996) and the xallarap event studied by Alcock et
al. (2001a).  In both cases, the lens preferentially lies in the
Magellanic Clouds. In addition, there has been the direct imaging of
another of the microlenses by Alcock et al. (2001b), revealing it to
be a nearby low-mass star in the disk of the Milky Way.

The claims of Belokurov et al. (2003, 2004) and especially Evans \&
Belokurov (2004) were challenged by Griest \& Thomas (2005).  Of
course, there is no need to re-enact the epic battle between the mice
and the frogs (Homer, 8th century BC) in the pages of this {\it
Journal}. Nonetheless, Griest \& Thomas (2005) did make a number of
scientifically incorrect statements regarding our neural network
computations. The main aim of this paper is simply to set the record
straight with regard to the event selection (\S2) and the efficiency
calculation (\S3).  In our discussion (\S4), we delineate the
remaining causes of scientific disagreement and discuss ongoing
experiments that may provide a resolution.

\begin{figure*}
\includegraphics[width=10cm]{\FigDir/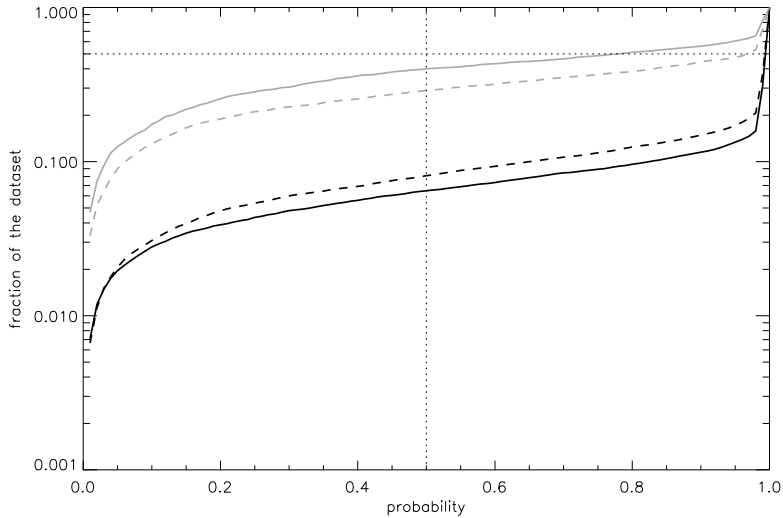}
\caption{The $\chi^2$ value of the constant baseline model is computed
for each lightcurve using (i) the publicly available data from the
MACHO website (black curves) and (ii) the cleaned data (grey curves)
used in Alcock et al's (2000) analysis.  The full and dashed curves
are the empirical cumulative probability distributions $P(\chi <
\chi_0)$ in the red and blue passbands respectively.  The vertical
axis is the fraction of the dataset. The horizontal axis would
normally be the chi-squared value $\chi_0$ in a cumulative probability
distribution. However, if the errors are normally distibuted, the
theoretical cumulative probablity distribution is known and is the
incomplete gamma function $P = \Gamma (0.5,\chi_0/2)$. This is plotted
as the horizontal axis. If the noise is Gaussian, then the probability
distribution passes through the point ($0.5,0.5$), whose location is
marked by the dotted lines. As the cleaned data pass close to this
point, we conclude that their noise properties are nearly Gaussian. By
contrast, the noise properties of the publicly available data are not
close to Gaussian at all.}
\label{fig:noisy}
\end{figure*}

\section{Remarks on Event Selection}

Any treatment of this subject should begin with some humbling
remarks. Neither Griest \& Thomas (2005) with powerful statistical
methods nor Belokurov et al. (2004) with neural networks can really
claim to have devised methods for microlensing detection that are
completely successful.  A striking indication of this is provided by
the EROS collaboration's discovery that the event MACHO LMC-23 is a
variable star (Tisserand \& Milsztajn 2005). The lightcurve for this
event is a good fit to a blended microlensing curve. Alcock et
al. (2000) report a $\chi^2$ of $1.452$ per degree of freedom. The
event was included in their set of confident microlensing events
(``set A''). Likewise, Belokurov et al. (2004) assessed the
probability of microlensing as $P =0.99$.  Therefore, both methods
failed.

The implications of this for microlensing surveys are worrisome.
There exist classes of variable stars whose lightcurves are good fits
to blended microlensing. They cannot be distinguished from
microlensing, except by more accurate photometric measurements or by
long-baseline monitoring for repeat variations. 


Belokurov et al. (2003, 2004) pioneered the use of neural networks to
identify microlensing by single lenses. Our calculations showed that
-- using the publicly available data -- only 7 of the events of
Alcock et al. (2000) are consistent with blended microlensing and
added Gaussian noise. These calculations are correct, but the noise in
the actual experiment is more complicated than Gaussian.

Notice that the selection of events by Alcock et al. (2000) makes the
same assumption that the noise is close to Gaussian, in order to
proceed with lightcurve fitting and the use of the $\chi^2$
statistic. However, the data through which Alcock et al. (2000) search
for events is not the publicly available data, but is derived from the
publicly available data by a cleaning process (see Alcock et
al. 1997). We refer to this as the cleaned dataset; it is not publicly
available.

Let us compare the noise properties of the public data with the
cleaned data.~\footnote{One tile only (roughly 3000 lightcurves) of
the cleaned data was made available to us for comparison purposes. We
thank Andrew Drake for making this possible The public but uncleaned
data are available at
http://wwwmacho.mcmaster.ca/Data/MachoData.html.} To estimate the
amount of variability in the lightcurve, we calculate the $\chi^2$
value of the constant baseline model. The empirical cumulative
probability distributions $P(\chi < \chi_0)$ are then constructed and
shown in Figure~\ref{fig:noisy} as full and dotted curves. The
vertical axis is the fraction of the dataset (and so runs from 0 to
1). In a typical cumulative probability distribution, the horizontal
axis would be the chi-squared value, $\chi_0$. However, here we have
converted this to a probability using the fact that, if the errors are
normally distributed, the theoretical cumulative probability
distribution is known and is the incomplete gamma function
$\Gamma(0.5, \chi_0/2)$ (see Press et al. 1992). In
Figure~\ref{fig:noisy}, the cumulative probability distributions in
the full black and dashed black lines refer to the public data in the
MACHO instrumental $R$ and $B$ filters. The full and dashed grey lines
refer to the cleaned data in Johnson $V$ and Kron-Cousins $R$ (see
Alcock et al. 1997). We see immediately that the noise properties of
the two datasets are very different. In the ideal case of Gaussian
noise and non-varying lightcurves, the probability distributions
should have a similar shape but pass through the point ($0.5,
0.5$). For the public data, $\sim 90 \%$ of the lightcurves correspond
to varying objects. By contrast, the noise properties of the cleaned
data are much closer to Gaussian.

\begin{figure*}
\includegraphics[height=16cm,width=18cm]{\FigDir/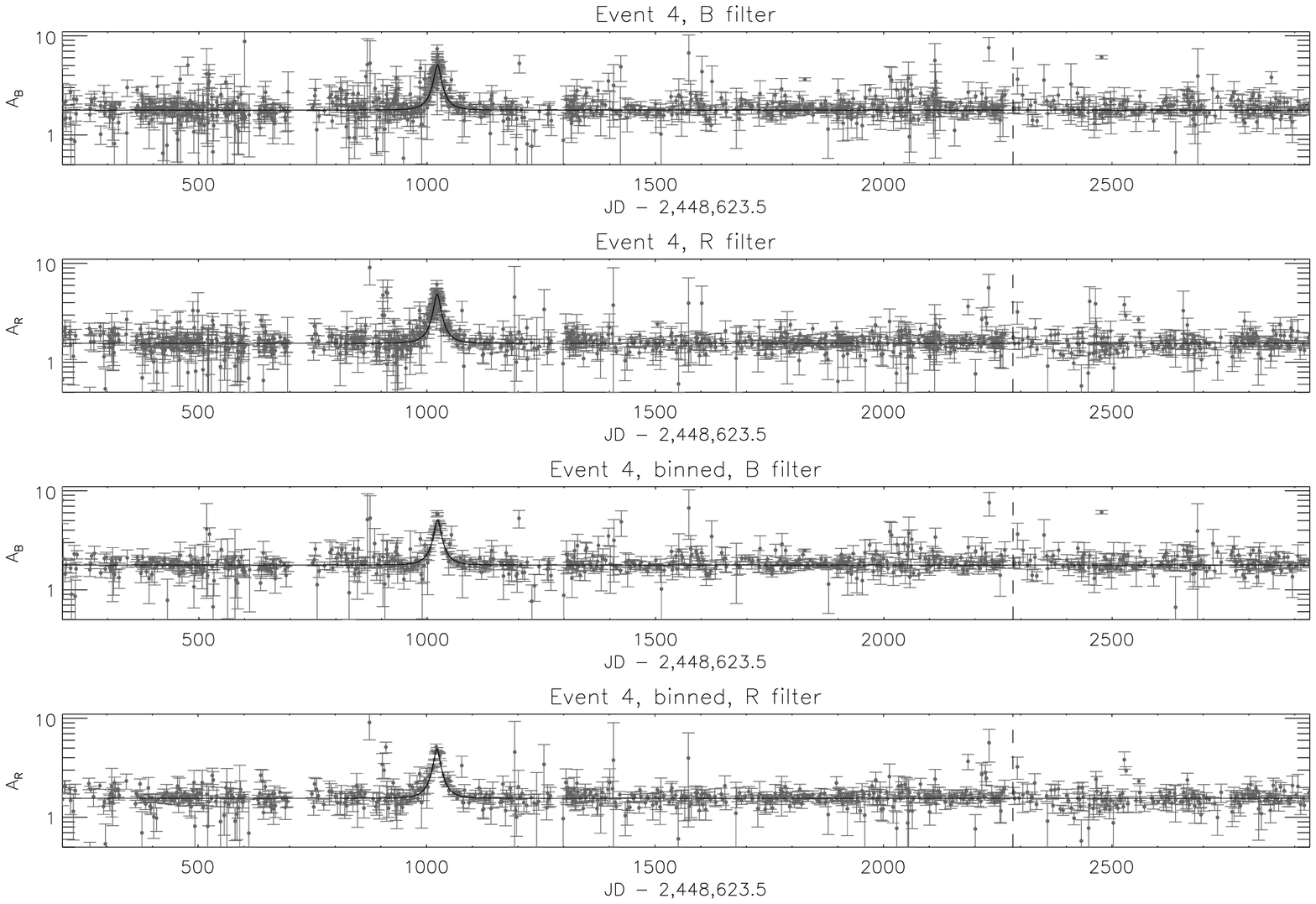}
\caption{The publicly available $R$ and $B$ band data for the event
MACHO LMC-4. The upper two panels show the unbinned data, the lower
two panels the binned data (with bin-size 1 day). The horizontal axis
is time in JD - 2448623.5 (2 Jan 1992), whilst the vertical axis is
the linear magnification $A_{R,B}$. The peak of the event, which is
discernible from the microlensing fit superposed on the lightcurves,
is at 1023.0 days and the Einstein crossing time $t_E$ is 43
days. The corresponding lightcurves using the cleaned data are
displayed in Alcock et al. (2000).}
\label{fig:publiclc4}
\end{figure*}
%

Let us illustrate the points at issue with an example.
Figure~\ref{fig:publiclc4} shows the $R$ and $B$ band data of MACHO
LMC-4, using the public data. The upper panels show the unbinned data,
the lower panels show the binned data. This figure should be compared
with the published version of the lightcurve of the same event in
Alcock et al. (2000). In particular, the lightcurve of
Figure~\ref{fig:publiclc4} shows secondary activity outside the
microlensing bump, for example, at $t \approx 1400$ days in the $B$
band.  These datapoints are not present in Alcock et al. (2000) and so
they must have been rejected at an early stage in the photometric
reduction. There may of course have been good reasons to remove these
datapoints. 

Griest \& Thomas (2005) argue that the greater number of lightcurves
identified as microlensing by Alcock et al. (2000) is due to the power
of lightcurve fitting. This is not the case. Lightcurve fitting will
not identify the data in Figure~\ref{fig:publiclc4} as a microlensing
event. To show this, we perform exactly the lightcurve fitting that
Griest \& Thomas (2005) advocate. The $\chi^2$ per degree of freedom
is 1.73 for the publicly available lightcurve of MACHO LMC-4. This is
higher than the value of 1.38 reported by Alcock et al. (2000). In
fact, there are no lightcurves in Alcock et al.'s (2000) set A with as
high a reduced $\chi^2$ as 1.73.

In other words, the main differences between the results of Belokurov
et al. (2004) and those of Griest \& Thomas (2005) are caused by the
fact that these investigators search through different datasets. More
succinctly, {\it the publicly available MACHO data are badly polluted
with photometric outliers.}

Of course, the MACHO collaboration has much more information on
photometric problems than is publicly available.  For example, they
can remove datapoints based on photometric problems that recur for
hundreds of stars in the same field for a troubling exposure (Griest
2005, private communication). At least as judged from
Figure~\ref{fig:noisy}, they seem to have done a good job. Hence, we
believe that the cleaned data are more useful than the public data --
it is just a pity that the cleaned data are not generally available.

Belokurov et al. (2004) realised that there are problems with the
public data and discarded any data-point that deviates more than $3
\sigma$ from its neighbours.  Griest \& Thomas (2005) suggest that
this procedure removes fast, short duration events and so is partly
responsible for the discrepancy.  For each lightcurve, Belokurov et
al. (2004) analyse {\it both} the public data and the data from which $3
\sigma$ outliers have been removed and quote the maximum output from
the neural networks (i.e., the one for which the probability of
microlensing is the greatest). Hence, if the public data has a very
high probability of microlensing, the event is recognized.

Griest \& Thomas (2005) assert that Belokurov et al.'s (2004) cross and
auto-correlations assume that the photometric data are evenly spaced
in time.  Neural networks are pattern recognition machines. So, they
can be trained to recognize the pattern of a sparsely sampled
microlensing lightcurve, without explicit accounting for missing
data. Of course, it is a different pattern to a microlensing
lightcurve with full time-sampling -- but a pattern none the less.
All that is needed is to include examples of such patterns with sparse
sampling in the training set, as Belokurov et al. (2004) do.  Besides,
one of the features input to the neural network, namely the
Lomb-Scargle periodogram, does not assume equally spaced datapoints
(see for example, Press et al. 1992).


Griest \& Thomas (2005) have also miscalculated the false positive
rate of the algorithm in Belokurov et al. (2004). The false positive
rates stated in Belokurov et al. (2004) refer to {\it the rates at
which the common classes of variable stars in the training set are
mistakenly identified as microlensing.} They do not refer to the rate
at which any lightcurve (whether variable or constant baseline) is
mistakenly identified as microlensing.  To compute the number of false
positives for the whole of the 9 million lightcurves in the MACHO
database, we must first calculate how many variable stars are
expected?  We can estimate this using results from the OGLE--II survey
of the LMC (\.{Z}ebru\'n et al. 2001).  OGLE-II found that $0.8 \%$ of
all stars towards the LMC are variables with amplitude variations
exceeding the measurement errors and with periods less than a few
years. The MACHO collaboration monitored $\sim 11.9$ million
lightcurves, of which $20 \%$ occur in field overlaps (Alcock et
al. 2000). So, this implies that the number of true variable stars is
$\sim 85 000$.  From the experiment reported in Figure 3 of Belokurov
et al. (2004), the false positive rate is 2 in 22000 or $0.9 \times
10^{-4}$.  Given that the total number of variable stars is $\sim 85 000$,
this means that the false positives amount to less than 8 for the
whole dataset of 9 million lightcurves.

However, the experiment reported in Figure 3 of Belokurov et
al. (2004) used the public data, not the cleaned data. As we have
shown in Figure~\ref{fig:noisy} here, this public dataset is very
noisy and contains over an order of magnitude too many variable
objects (mainly caused by artefacts). Hence, when applied to the
dataset that the MACHO collaboration actually use, our false positive
will certainly diminish, probably by at least an order of magnitude.

Griest \& Thomas (2005) point out that the MACHO selection procedure
uses around 20 statistical methods, whereas Evans \& Belokurov (2004)
use only 5 as inputs to the neural networks. They suggest that their
suite of statistical methods may be more powerful. In fact, it is
difficult to tell unless both methods are run on the same
data. However, as Figure 2 of Belokurov et al. (2004) shows, the 5
statistical methods already identify 95 \% of microlensing events in
the test set. This is an excellent result by any standards -- and it
is easy to incorporate any further statistics as inputs to the neural
networks, if desired.
 
Finally, Griest \& Thomas (2005) mistakenly argue that Evans \&
Belokurov (2004) only looked for microlensing events among the 22
MACHO candidates and so the efficiency is necessarily reduced. Let us
clarify the calculation that was actually done. Evans \& Belokurov
(2004) examined 22 000 lightcurves, including the 22 MACHO candidates.
This number includes the ``set A'' or stronger microlensing candidates
and the ``set B''or weaker candidates from Alcock et al. (2000).  We
made the additional, strong assumption that there are no further
microlensing events in the rest of the data. This assumption may be
valid or invalid -- it is impossible to say unless all the data are
made available for testing with neural networks\footnote{At present,
lightcurves can be downloaded on an individual basis (or in small
groups) from the MACHO website, thus making downloads of 12 million
lightcurves practically impossible. In any case, what is really needed
is the capability to make downloads of entire fields of the cleaned
data.}. However, there is some supporting evidence for Evans \&
Belokurov's (2004) point of view, as none of the ``set B'' MACHO
candidates passed our neural network selection criterion. Having made
this assumption, Evans \& Belokurov (2004) proceed -- as usual in any
efficiency calculation -- by calculating the fraction of simulated
events that are identified by the algorithm. Such an efficiency
calculation does not depend on whether all or part of the original
dataset was analyzed.

\section{Remarks on False Positives}

Any treatment of this subject should begin with some cautionary
remarks on the limitations of efficiency calculations.  The optical
depth is usually calculated from the data as a sum over events using
\begin{equation}
\tau = {\pi \over 4} \sum_i {t_{0,i} \over N T \epsilon(t_{0,i})}
\end{equation}
where $N$ is the number of stars monitored, $T$ is the duration of the
experiment and $\epsilon$ is the efficiency as a function of
timescale.  There is an important assumption in this formula.  The
assumption is that the false positive rate is completely
negligible. {\it An efficiency calculation can correct for false
negatives (that is, missed microlensing events) but it cannot correct
for false positives (that is, variable objects wrongly classified as
microlensing).}

By altering the threshold in any selection procedure, we simply
generate more false positives at the expense of less false negatives
-- or vice versa. Even after correcting with the efficiency, different
selection procedures will not give the same optical depth, unless the
false positive rate is completely negligible for all the
thresholds. Alcock et al. (1997) and Alcock et al. (2000) used
different cuts and obtained quite different values for the optical
depth. This is a broad hint that the false positive rate is not
negligible in the Alcock et al. (1997) sample.

Everyone now accepts that MACHO LMC-23 is a variable star (Tisserand
\& Milsztajn 2005). This false positive was included in the more
confident ``set A'' of Alcock et al. (2000). ``Set B'' has a lower
threshold and logically must contain still more false positives. In
any case, let us emphasise that there is nothing exotic about MACHO
LMC-23. It is a variable star that is able to masquerade as a
microlensing event because of photometric noise. There is already some
indication of this in the comparatively high value of its $\chi^2$ of
$1.452$ per degree of freedom.

Given the fact that there are false positives in the Alcock et al.
(2000) samples, what is the best methodology for correcting the
optical depth?  There have been three attempts to do this so far,
namely by Griest \& Thomas (2005), Bennett (2005) and Evans \&
Belokurov (2004). All three computations have inadequacies.

Griest \& Thomas (2005) attempted to correct for the effects of
contamination by removing the contribution of MACHO LMC-23. To see why
this inappropriate, let us consider a model problem in which the cut
is only based on a $\chi^2$ per degree of freedom. Then three further
events (candidates $5, 8$ and $21$) would be discarded, as their
$\chi^2$ per degree of freedom is worse than MACHO LMC-23. This is
because the decision boundary between microlensing and
non-microlensing cannot have artificially created holes, or any abrupt
or sharp features.  Of course, the actual algorithm that the MACHO
collaboration use is more sophisticated than a cut on $\chi^2$. Our
point is merely that a misclassified event inside the decision
boundary affects its entire neighbourhood. It is not enough simply to
remove by hand the contribution of the misclassified event, as Griest
\& Thomas (2005) do. Their calculation is not a proper accounting of
the effects of contamination, even in the optimistic case in which
MACHO LMC-23 is the only false positive.

Bennett (2005) tried to correct the optical depth calculations of
Alcock et al. (2000) for the effects of contamination by introducing a
likelihood estimator.  His likelihood estimator is tantamount to
assuming that the contamination rate is 1 event out of every 5.
Although Bennett (2005) does recompute the efficiencies, this is to take
into account a systematic error in the efficiencies used by Alcock et
al. (2000). However, Bennett does not take into account the change in
the efficiencies caused by the different event selection required to
eliminate the contaminants.

Evans \& Belokurov (2004) published an estimate of the optical depth
allowing for contamination. We argued for more contaminants than
either Bennett (2005) or Griest \& Thomas (2005) and concluded that
the optical depth $\tau$ satisfied $0.3 \times 10^{-7} \lta \tau \lta
0.5 \times 10^{-7}$, where the range corresponds to the $\pm 1 \sigma$
interval. Evans \& Belokurov's (2004) analysis has two problems.  First, as
stated earlier, we did not have access to the whole dataset, but to
$\lta 1 \%$ of it. Second, in our efficiency calculation, we used a
simple microlensing lightcurve model with added Gaussian noise and
MACHO sampling to test whether events would be found by neural
networks. This procedure would be better applied to the cleaned data,
which are not available, and so Evans \& Belokurov (2004) perforce
used the polluted public data.

If a decision boundary between microlensing and non-microlensing is
introduced, then it is crucial to know the false positive rate. We
have not been able to find such a calculation in the literature for
the MACHO experiment. To compute the false positive rate, it is
important to use the full gamut of possible variable star lightcurves.
In Belokurov et al. (2004), the false positive rate was calculated
using standard libraries of variable stars.  Another possibility is to
apply Feeney et al.'s (2005) adaption of the technique of $K$ fold
cross-validation, which uses the entire dataset itself to provide the
range of variable lightcurves.

The alternative to introducing a decision boundary is to assign
probabilities to each lightcurve using, for example, the outputs of
neural networks.  The microlensing rate can be calculated directly
from the outputs, without introducing an explicit decision
boundary. Every lightcurve makes a weighted contribution to the
microlensing rate.

One way of carrying this out is described in Belokurov et al. (2004,
see Appendix A). Briefly, the formula
\begin{equation}
{\hat P} ({\rm microlensing}) \approx \frac{1}{N}\sum_{i} y_i
\end{equation}
is used to estimate the true probability of microlensing.  Here, $i$
runs through all $N$ lightcurves in the entire data set.  The
probabilities $y_i$ are the outputs of neural networks. Ordinarily,
the output is the posterior probability of microlensing, given the
prior probabilities imposed by the training set. However, Belokurov
et al. (2004) showed how to iteratively adjust the outputs so that
they converged to the true probabilities given the real-world priors.

The output of this procedure is the probability of microlensing in the
experiment monitoring $N$ stars and lasting for a duration $T$. From
this, the microlensing rate is
\begin{equation}
\Gamma = {N \over T}{\hat  P}({\rm microlensing})
\end{equation}
The advantage of this algorithm is that the rate can be computed
directly from the dataset, without the intervening steps of candidate
selection and efficiency estimation.

This is very different to the approach of all microlensing experiments
so far, which have categorized events as either microlensing or
non-microlensing. The probabilities assigned are therefore either $1$
or $0$. Not merely are marginal events incorporated into the optical
depth with the same weight as unambiguous events, but -- worse still
-- their contribution is amplified by the efficiency factor as well.
The efficiency naturally tends to be low for the marginal events.
This may well be part of the reason for the continuing mismatch
between theoretical estimates and observational results in
microlensing.

\begin{table*}
\caption{The theoretical value of the microlensing optical depth
towards the LMC for various known stellar populations in the Milky Way
and the Large Magellanic Clouds.}
\label{table:ods}
\begin{tabular}{lcc}
\hline
Component & Optical Depth & Notes \\
\hline
Thin and Thick disk    & $0.10 \times 10^{-7}$ & 
Eqn.(2) of Binney \& Evans (2001) using a local column density of \\
\null & \null & $27 {\rm M}_\odot\,{\rm pc}^{-2}$ and a radial scalelength of
3.0 kpc \\
Spheroid     & $0.02 \times 10^{-7}$ & Standard $\rho = 1.18 \times
10^{-4} (r/R_0)^{-3.5}{\rm M}_\odot\,{\rm pc}^{-3}$ spheroid of 
Giudice et al. (1994) \\
LMC disk/bar & $0.55 \times 10^{-7}$ & Zero Offset Model 
of Zhao \& Evans (2001)\\
LMC disk/bar & $1.0 \times 10^{-7}$ & Non-Zero Offset Model 
of Zhao \& Evans (2001)\\
LMC disk/bar & $0.05 - 0.80 \times 10 ^{-7}$ & Models of Gyuk, Dalal
\& Griest (2000)\\
\hline
\end{tabular}
\end{table*}

\section{Discussion} 

Evans \& Belokurov's (2004) suggestion that the optical depth to the
LMC may have been over-estimated because of contamination by false
positives deserves serious consideration.  Indeed, the suggestion
receives support from the subsequent results of the EROS
collaboration, which was also monitoring the Large Magellanic
Cloud. Tisserand \& Milsztajn (2005) find the low optical depth $\tau
= (0.15 \pm 0.12) \times 10^{-7}$ based on 6.7 years of data. They also
report that MACHO LMC-23 -- included in Alcock et al.'s (2000) ``set
A'' of confident events -- is actually a variable star. Very recently,
Tisserand et al. (2006) exploited the idea of a bright subsample to
minimise the effects of contamination and obtained $\tau < 0.38 \times
10^{-7}$ 

The main reason for the difference between the results of Belokurov et
al. (2004) and those of Griest \& Thomas (2005) lies in the treatment
of the photometric outliers.  Belokurov et al.'s (2004) calculations
use the public data and are valid in the case that the noise is
Gaussian.  In fact, the noise properties of the public data are
non-Gaussian. Griest \& Thomas (2005) use the cleaned data -- a
version of the data in which many photometric outliers have been
removed -- so that the noise is closer to Gaussian. The cleaned data
are not publicly available. The events that Belokurov et al. (2004)
claimed as microlensing are reasonably trustworthy. If an event is
identified in noisy data, then use of the cleaned data will only
improve matters. More problematic are the events for which Belokurov
et al. (2004) failed to identify as microlensing, at variance with the
original judgement of Alcock et al. (2000).  It is impossible to say
anything further about these events until either the cleaned data or
the algorithm for cleaning the public data are made public. Belokurov
et al. (2004) therefore give a final sample of events whose
microlensing nature is almost beyond question. This is valuable, as
false positives are destructive and cannot be corrected by the
efficiency.

Let us also remark that the events Belokurov et al. (2004) claimed as
non-microlensing {\it may} be incorrectly designated (c.f. Bennett,
Becker \& Tomaney 2005). If so, this is not a fault of the neural
network methods, but a consequence of the use of the polluted public
data.

Are Griest \& Thomas (2005) correct to claim a microlensing puzzle?  It is
true that the experimental determinations of the optical depth to the
LMC are presently uncertain to almost an order of magnitude (from
$\tau = (0.15 \pm 0.12) \times 10^{-7}$ based on EROS data by Tisserand
\& Milsztajn (2005) to $(1.0\pm 0.3) \times 10^{-7}$ based on MACHO data
by Bennett (2005)). However, the EROS experiment monitors a wider
solid angle of less crowded fields in the LMC than the MACHO
experiment. So, blending and contamination by LMC self-lensing are
less important for the EROS experiment than for MACHO. The EROS result
is therefore an average value of the optical depth over a wide area of
the LMC disk, whilst the MACHO value is the optical depth in the
central parts. Nonetheless, this cannot be the whole story. The
contribution to the optical depth of lensing objects lying in the
Milky Way halo varies only weakly across the face of the LMC. So, if
the claims that $20 \%$ of the dark halo is in the form of compact
objects are correct (e.g., Alcock et al. 2000; Griest \& Thomas 2005),
then this optical depth contribution of this lensing population
(approximately $\tau \sim 0.6 \times 10^{-7}$) should be largely
independent of position.

The theoretical estimates of the optical depth of the known Galactic
components in the direction of the LMC have been computed anew and are
listed in Table~\ref{table:ods}. Using the latest models of the thin
and thick disk (e.g., Binney \& Evans 2001), we find that their
contribution is $\tau = 0.10 \times 10^{-7}$. This is a
middle-of-the-range value, and both larger (e.g., Alcock et al. 1997;
Evans et al 1998) and smaller numbers (Alcock et al. 2000) can be
found in the literature. The optical depth of the spheroid is
uncontroversial and is $\tau = 0.02 \times 10^{-7}$. There is much
more dispute about the LMC self-lensing optical depth. Accordingly, we
list a number of recent estimates in the Table -- our preferred value
is $0.55 \times 10^{-7}$, corresponding to the zero offset model of
Zhao \& Evans (2001), which is again a reasonable middle-of-the-range
value. Notice from Figure 2 of Zhao \& Evans (2001) that the LMC
optical depth is roughly constant over the central 2 kpc of the LMC
bar.  Adding these numbers up, the total optical depth contribution to
the LMC from known sources is $0.67 \times 10^{-7}$. The error on this
theoretical estimate is large, as it controlled by the poorly known
extension along the line of sight of the LMC.

The most recent experimental determination of the optical depth from a
MACHO collaboration member now stands at $\tau = (1.0 \pm 0.3) \times
10^{-7}$ (Bennett 2005). In our judgement, Bennett's calculation is an
overestimate. Nonetheless, even accepting his value, the experimental
optical depth is only just over $1\sigma$ away from the theoretical
estimate from known populations.  There is no major puzzle! This is
especially the case given the uncertainties in the physical depth of the
LMC. For example, Weinberg \& Nikolaev (2000) detected a spread of a
few kiloparsecs in distance among their 2 Micron All-Sky Survey of LMC
stars. If this is a true indication of the line of sight depth, then
the LMC optical depth is still higher than we have assumed, and even
the remaining small discrepancy melts away.

There are a number of ongoing experiments that may help solve the
microlensing puzzle in the near future. First, and most promisingly,
the super-MACHO survey (Stubbs 1999, Becker et al. 2004) has the
specific goal of identifying the location of the objects producing the
microlensing events. The survey has been taking data since 2001 and
has extensive coverage of the face of the LMC. If the lensing objects
lie in the halo, there is only a weak gradient across the face of the
LMC. However, if the objects lie in the LMC, then there is a
substantial gradient. The measurement of this gradient requires an
order of magnitude more events than those reported by MACHO and EROS,
but should be within the grasp of super-MACHO.

Second, there are a number of microlensing experiments towards the
Andromeda galaxy, such as the POINT-AGAPE and MEGA experiments (e.g.,
Paulin-Henriksson et al. 2003; de Jong et al. 2004; Belokurov et
al. 2005), that are now reporting results. These experiments are
motivated by the suggestion of Crotts (1992) that the event rate to
sources in the near and far disks in M31 is different. The lines of
sight to the far disk as compared to the near disk are longer and pass
through more of the M31 dark halo.  However, An et al. (2004) showed
that the expected microlensing asymmetry between the near and far disk
is overwhelmed by the effects of patchy and variable extinction in the
M31 disk. Of course, it is considerably harder to provide convincing
evidence that a claimed pixel-lensing event in M31 is due to
microlensing as compared to resolved microlensing events. Perhaps
because of this, the independent calculations of the optical depth of
the M31 halo currently in the literature are contradictory. Calchi
Novati et al. (2005) found 6 short duration events and argued that at
least 20 per cent of the M31 halo is in the form of dark, compact
objects. De Jong et al. (2006) identified 14 events, but concluded
that the signal was equally consistent with both self-lensing and with
dark, compact halo objects.

\section*{Acknowledgements}
We thank Piotr Popowski and Kim Griest for critical readings of the
manuscript, as well as the anonymous referee for many helpful
suggestions. VB acknowledges financial support from the Particle
Physics and Astronomy Research Council of the United Kingdom.

{}


\begin{thebibliography}{}

\bibitem[Alcock et al.(1997)]{1997ApJ...479..119A} Alcock C.~et al.\ 1997, 
ApJ, 479, 119 

\bibitem[Alcock et al.(2000)]{2000ApJ...542..281A} Alcock C.~et al.\ 2000, 
ApJ, 542, 281 

\bibitem[Alcock et al.(2001a)]{2001a} Alcock C.~et al.\ 2001a, 
ApJ, 552, 259 

\bibitem[Alcock et al.(2001b)]{al2001b} Alcock C., et al.\ 2001b, 
Nature, 414, 617
 
\bibitem[An et al.(2004)]{an04b}
An J.H. et al.\ 2004, MNRAS, 351, 1071
 
\bibitem[Becker et al.(2004)]{Be04}
Becker A., et al. 2004, in ``Proceeding of IAU Symposium 225: Impact of
Gravitational Lensing on Cosmology'', eds, Y. Mellier, G. Meylan, p. 357
(Cambridge University Press, Cambridge), astro-ph/0409167

\bibitem[Belokurov, Evans, \& Du(2003)]{bela} Belokurov V., 
Evans N.W., Le Du Y. 2003, MNRAS, 341, 1373 

\bibitem[Belokurov, Evans, \& Du(2004)]{bel1} Belokurov V., 
Evans N.W., Le Du Y. 2004, MNRAS, 352, 233 

\bibitem[Belokurov, Evans, \& Du(2004)]{bel2} Belokurov V., et
  al. 2005, MNRAS, 357, 17

\bibitem[Bennett (1996)]{bennetta} Bennett D. 1996, Nuc. Phys. B,
51B, 131

\bibitem[Bennett (2005)]{bennettb} Bennett D. 2005, ApJ, 633, 906

\bibitem[Bennett et al. (2005)]{bennettc} Bennett D., Becker A., Tomaney
  A. 2005, ApJ, 631, 301

\bibitem[Binney \& Evans (2001)]{be}
Binney J.J., Evans N.W. 2001, MNRAS, 327, L27

\bibitem[Calchi Novati (2005)]{cn}
Calchi Novati et al. S. 2005, A\&A, 443, 911

\bibitem[Evans et al. 1998)]{evans1} Evans N.W., Gyuk G.,
Turner M.S., Binney J.J. 1998, ApJ, 501, L45

\bibitem[Evans(2004)]{evans2} Evans N.W., Belokurov V. 2004, In
``IDM:2002 Fifth International Conference on the Identification of
Dark Matter'', eds N. Spooner, V. Kudryavtsev, p. 141 (World
Scientific, Singapore), astro-ph/0411222

\bibitem[Feeney et al. (2005)]{sf}
Feeney S. et al. 2005, AJ, 130, 84

\bibitem[Giudice et al. (1994)]{gian}
Giudice G.F., Mollerach S., Roulet E., Phys. Rev D, 50, 2406

\bibitem[Griest \& Thomas(2005)]{gt}
Griest K., Thomas C. 2005, MNRAS, 359, 464

\bibitem[Gyuk, Dalal \& Griest (2000)]{gdg}
Gyuk G., Dalal N., Griest K. 2000, ApJ, 535, 90

\bibitem[Han \& Gould(1995)]{han}
Han C., Gould A. 1995, ApJ, 449, 521

\bibitem[Homer]{homer} Homer, 1914, Batrachomyomachia,  Loeb Classical
Library 57 (Heinemann: London)

\bibitem[de Jong et al.2004]{jelte1}
De Jong J., et al. 2004, AA, 417, 461

\bibitem[de Jong et al.2006]{jelte2}
De Jong J., et al. 2006, AA, 446, 855

\bibitem[Kerins et al.(2004)]{ek1}
Kerins E.J. 2004, MNRAS, 347, 1033

\bibitem[Lasserre et al.(2000)]{2000A&A...355L..39L} Lasserre T.~et al.\ 
2000, AA, 355, L39 

\bibitem[Paulin-Henriksson et al.(2003)]{PH03}
Paulin-Henriksson S. et al., 2003, A\&A, 405, 15

\bibitem[Press et al. 1992]{pft} Press W.H., Teukolsky S., Vetterling
W.T., Flannery B. 1992. Numerical Recipes, (Cambridge University
Press: Cambridge)

\bibitem[Stubbs 1999]{stubbs} Stubbs C., 1999, In ``The Third Stromlo
Symposium: The Galactic Halo'', eds. B.K. Gibson, T.S. Axelrod,
M.E. Putman, ASP Conference Series 165, p. 503

\bibitem[Tisserand \& Milsztajn (2005)]{tm}
Tisserand P., Milsztajn A. 2005, Proceedings of 5th Rencontres du
Vietnam'', astro-ph/0501584

\bibitem]Tisserand et al.(2006)]{Ti06}
Tisserand P., 2006, AA, in press (astro-ph/0607207)

\bibitem[Weinberg \& Nikolaev (2000)]{wn}
Weinberg M., Nikolaev S. 2000, ApJ, 542, 804

\bibitem[Zebrun et al.(2001)]{zebrun}
\.{Z}ebru\'n K. et al. 2001, Acta Astron., 51, 317.

\bibitem[Zhao \& Evans (2000)]{ze}
Zhao H.S., Evans N.W. 2001, ApJ, 545, L35

\end{thebibliography}
\end{document}